\documentclass[a4paper]{jpconf}

\usepackage{graphicx}

\usepackage{comment}

\usepackage{hyperref}
\hypersetup{colorlinks=true, citecolor=blue, linkcolor=blue, urlcolor=blue, pdfborder={0 0 0}, breaklinks=true} 
\renewcommand{\url}[1]{\href{#1}{#1}}
\newcommand{\doi}[1]{\url{https://doi.org/#1}}
\newcommand{\arxiv}[1]{\href{https://arxiv.org/abs/#1}{arxiv:#1}}

\usepackage{multirow}
\usepackage{datetime2}

\usepackage[absolute,overlay]{textpos}

\usepackage{lineno}

\begin{document}

\title{Speeding up Madgraph5\_aMC@NLO
through\\
CPU vectorization and GPU offloading:\\
towards a first alpha release}

\author{A~Valassi$^1$,
T~Childers$^2$,
L~Field$^1$,
S~Hageb\"ock$^1$,
W~Hopkins$^2$,
O~Mattelaer$^3$,
N~Nichols$^2$,
S~Roiser$^1$,
D~Smith$^1$,
J~Teig$^1$,
C~Vuosalo$^4$,
Z~Wettersten$^1$}

\address{$^1$ IT Department, CERN, Geneva, Switzerland}
\address{$^2$ Argonne National Laboratory, USA}
\address{$^3$ Universit\'e Catholique de Louvain, Belgium}
\address{$^4$ University of Wisconsin-Madison, USA}

\ead{andrea.valassi@cern.ch}

\begin{textblock*}{18cm}(1cm,27.9cm) 
{\bf Version 2.0 (8 December 2023)}
\end{textblock*}
\begin{textblock*}{18cm}(1cm,28.35cm) 
\footnotesize{Minor updates over version 1.0 (31 March 2023)}
\end{textblock*}

\begin{abstract}
The matrix element (ME) calculation in any Monte Carlo physics~event~generator~is
an ideal fit for implementing data parallelism 
with lockstep processing 
on GPUs and vector CPUs.
For complex physics processes where the ME 
calculation~is~the~computational 
bottleneck of event generation workflows, this can lead to 
large overall speedups by efficiently exploiting these hardware architectures, which are now largely underutilized in HEP.~In~this 
paper,
we present the status of our work on the reengineering of the Madgraph5\_aMC@NLO~event generator 
at the time of the ACAT2022
conference.
The progress achieved since our previous
publication in the ICHEP2022 proceedings~\cite{bib:ichep2022} 
is discussed, for our implementations
of the ME calculations in vectorized C++,
in CUDA and in the SYCL framework,
as well as in their integration into the
existing MadEvent framework.
The outlook towards a first alpha release of the software supporting QCD LO processes 
usable by the LHC experiments
is also discussed.
\end{abstract}

\newcommand{\eemumu}{\mbox{$e^+\!e^-\!\!\rightarrow\!\mu^+\!\mu^-$}}
\newcommand{\ggtt}{\mbox{$gg\!\rightarrow\! t\bar{t}$}}
\newcommand{\ggttg}{\mbox{$gg\!\rightarrow\! t\bar{t}g$}}
\newcommand{\ggttgg}{\mbox{$gg\!\rightarrow\! t\bar{t}gg$}}
\newcommand{\ggttggg}{\mbox{$gg\!\rightarrow\! t\bar{t}ggg$}}

\newcommand{\muc}[1]{\multicolumn{2}{c|}{#1}}
\newcommand{\mur}[1]{\multirow{2}{*}{#1}}

\newcommand{\tabtputcudanew}[3]{
\begin{table}[#1]
\vspace*{#2} 
\begin{center}{
\small
\hspace*{-4mm}
\setlength\tabcolsep{5pt} 
\begin{tabular}{|l|c|c|c|c|c|c|}
\cline{3-7}
\multicolumn{2}{c|}{}& 
\multicolumn{3}{c|}{madevent}& 
\multicolumn{2}{c|}{standalone}\\
\hline
\multicolumn{2}{|c|}{CUDA grid size}& 
\multicolumn{4}{c|}{8192}& 
\multicolumn{1}{c|}{524288}\\
\cline{1-7}
\multirow{2}{*}{\ggttgg} &
MEs &
$t_\mathrm{TOT} = t_\mathrm{Mad} + t_\mathrm{MEs}$ &
$N_\mathrm{events} / t_\mathrm{TOT}$ &
\multicolumn{3}{c|}{$N_\mathrm{events} / t_\mathrm{MEs}$}\\
& precision &
[sec] &
[events/sec] &
\multicolumn{3}{c|}{[MEs/sec]}\\
\hline
Fortran & double &
55.4 = 2.4 + 53.0 &
1.63E3 (=1.0) &
1.70E3 (=1.0) &
--- &
--- \\
\hline
CUDA & double &
\hphantom{0}2.9 = 2.6 + 0.35 &
3.06E4 (x18.8) &
2.60E5 (x152) &
2.62E5 & 
4.21E5 (x247) \\ 
\hline
CUDA & float &
\hphantom{0}2.8 = 2.6 + 0.24 &
3.24E4 (x19.9) &
3.83E5 (x225) &
3.96E5 & 
8.77E5 (x516) \\ 
\hline
\end{tabular}
}\end{center}
\vspace*{-5mm}
\caption{
Processing times and throughputs for
90112 \ggttgg\ weighted events.~One 
core
of a CERN VM
(Intel Silver 4216 CPUs,
one NVidia V100 GPU),
cuda11.7 and gcc11.2 builds.
See Ref.~\cite{bib:ichep2022}
for further details,
e.g.~on
the difference between the
madevent and standalone columns.}
\label{tab-tputcudanew}
\vspace*{#3} 
\end{table}
}

\newcommand{\tabtputcppnew}[3]{
\begin{table}[#1]
\vspace*{#2} 
\begin{center}{
\small
\hspace*{-4mm}
\setlength\tabcolsep{5pt} 
\begin{tabular}{|l|c|c|c|c|c|}
\cline{3-6}
\multicolumn{2}{c|}{}& 
\multicolumn{3}{c|}{madevent}& 
\multicolumn{1}{c|}{standalone}\\
\hline
\multirow{2}{*}{\ggttgg} & MEs &
$t_\mathrm{TOT} = t_\mathrm{Mad} + t_\mathrm{MEs}$ &
$N_\mathrm{events} / t_\mathrm{TOT}$ &
\multicolumn{2}{c|}{$N_\mathrm{events} / t_\mathrm{MEs}$}\\
& precision &
[sec] &
[events/sec] &
\multicolumn{2}{c|}{[MEs/sec]}\\
\hline
Fortran(scalar) & double &
37.3 = 1.7 + 35.6 &
2.20E3 (=1.0) &
2.30E3 (=1.0) &
--- \\
\hline
C++/none(scalar) & double&
37.8 = 1.7 + 36.0 &
2.17E3 (x1.0) &
2.28E3 (x1.0) &
2.37E3 \\
C++/sse4(128-bit) & double &
19.4 = 1.7 + 17.8 &
4.22E3 (x1.9) &
4.62E3 (x2.0) &
4.75E3 \\
C++/avx2(256-bit) & double &
\hphantom{0}9.5 = 1.7 + \hphantom{0}7.8 &
8.63E3 (x3.9) &
1.05E4 (x4.6) &
1.09E4 \\
C++/512y(256-bit) & double &
\hphantom{0}8.9 = 1.8 + \hphantom{0}7.1 &
9.29E3 (x4.2) &
1.16E4 (x5.0) &
1.20E4 \\
C++/512z(512-bit) & double &
\hphantom{0}6.1 = 1.8 + \hphantom{0}4.3 &
1.35E4 (x6.1) &
1.91E4 (x8.3) &
2.06E4 \\
\hline
C++/none(scalar) & float &
36.6 = 1.8 + 34.9 &
2.24E3 (x1.0) &
2.35E3 (x1.0) &
2.45E3 \\
C++/sse4(128-bit) & float&
10.6 = 1.7 + \hphantom{0}8.9 &
7.76E3 (x3.6) &
9.28E3 (x4.1) &
9.21E3 \\
C++/avx2(256-bit) & float&
\hphantom{0}5.7 = 1.8 + \hphantom{0}3.9 &
1.44E4 (x6.6) &
2.09E4 (x9.1) &
2.13E4 \\
C++/512y(256-bit) & float&
\hphantom{0}5.3 = 1.8 + \hphantom{0}3.6 &
1.54E4 (x7.0) &
2.30E4 (x10.0) & 
2.43E4 \\
C++/512z(512-bit) & float&
\hphantom{0}3.9 = 1.8 + \hphantom{0}2.1 &
2.10E4 (x9.6) &
3.92E4 (x17.1) & 
3.77E4 \\
\hline
\end{tabular}
}\end{center}
\vspace*{-5mm}
\caption{
Processing times and throughputs 
for 
81952 \ggttgg\ weighted events.
One core
of Juwels Cluster login node jwlogin07
(Intel Gold 6148 CPUs), 
gcc11.2 builds.
See Ref.~\cite{bib:ichep2022}
for further details,
e.g.~on
the five different
vectorization levels
(none, sse4, avx2, 512y, 512z).}
\label{tab-tputcppnew}
\vspace*{#3}
\end{table}
}

\newcommand{\tabtputcudanewg}[3]{
\begin{table}[#1]
\vspace*{#2}
\begin{center}{
\small
\hspace*{-4mm}
\setlength\tabcolsep{5pt} 
\begin{tabular}{|l|c|c|c|c|c|c|}
\cline{3-7}
\multicolumn{2}{c|}{}& 
\multicolumn{3}{c|}{madevent}& 
\multicolumn{2}{c|}{standalone}\\
\hline
\multicolumn{2}{|c|}{CUDA grid size}& 
\multicolumn{4}{c|}{8192}& 
\multicolumn{1}{c|}{16384}\\
\cline{1-7}
\multirow{2}{*}{\ggttggg} &
MEs &
$t_\mathrm{TOT} = t_\mathrm{Mad} + t_\mathrm{MEs}$ &
$N_\mathrm{events} / t_\mathrm{TOT}$ &
\multicolumn{3}{c|}{$N_\mathrm{events} / t_\mathrm{MEs}$}\\
& precision &
[sec] &
[events/sec] &
\multicolumn{3}{c|}{[MEs/sec]}\\
\hline
Fortran & double &
1228.2 = 5.0 + 1223.2 &
7.34E1 (=1.0) &
7.37E1 (=1.0) &
--- &
--- \\
\hline
CUDA & double &
\hphantom{00}19.6 = 7.4 + \hphantom{00}12.1 &
4.61E3 (x63) &
7.44E3 (x100) &
9.10E3 & 
9.51E3 (x129) \\
\hline
CUDA & float &
\hphantom{00}11.7 = 6.2 + \hphantom{000}5.4 &
7.73E3 (x105) &
1.66E4 (x224) &
1.68E4 & 
2.41E4 (x326) \\
\hline
CUDA & mixed &
\hphantom{00}16.5 = 7.0 + \hphantom{000}9.6 &
5.45E3 (x74) &
9.43E3 (x128) &
1.10E4 & 
1.19E4 (x161) \\
\hline
\end{tabular}
}\end{center}
\vspace*{-5mm}
\caption{
Processing times and throughputs 
for 
90112 \ggttggg\ weighted events.
One core
of a CERN VM
(Intel Silver 4216 CPUs,
one NVidia V100 GPU), 
cuda11.7 and gcc11.2 builds.}
\label{tab-tputcudanewg}
\vspace*{#3}
\end{table}
}

\newcommand{\tabtputcppnewg}[3]{
\begin{table}[#1]
\vspace*{#2}
\begin{center}{
\small
\hspace*{-4mm}
\setlength\tabcolsep{5pt} 
\begin{tabular}{|l|c|c|c|c|c|}
\cline{3-6}
\multicolumn{2}{c|}{}& 
\multicolumn{3}{c|}{madevent}& 
\multicolumn{1}{c|}{standalone}\\
\hline
\multirow{2}{*}{\ggttggg} & MEs &
$t_\mathrm{TOT} = t_\mathrm{Mad} + t_\mathrm{MEs}$ &
$N_\mathrm{events} / t_\mathrm{TOT}$ &
\multicolumn{2}{c|}{$N_\mathrm{events} / t_\mathrm{MEs}$}\\
& precision &
[sec] &
[events/sec] &
\multicolumn{2}{c|}{[MEs/sec]}\\
\hline
Fortran(scalar) & double &
813.2 = 3.7 + 809.6 &
1.01E2 (=1.0) &
1.01E2 (=1.0) &
--- \\
\hline
C++/none(scalar) & double&
986.0 = 4.3 + 981.7 &
8.31E1 (x0.8) &
8.35E1 (x0.8) &
9.82E1 \\
C++/sse4(128-bit) & double &
514.7 = 4.2 + 510.5 &
1.59E2 (x1.6) &
1.61E2 (x1.6) &
1.95E2 \\
C++/avx2(256-bit) & double &
231.6 = 4.0 + 227.6 &
3.54E2 (x3.5) &
3.60E2 (x3.6) &
4.41E2 \\
C++/512y(256-bit) & double &
208.6 = 3.9 + 204.8 &
3.93E2 (x3.9) &
4.00E2 (x4.0) &
4.95E2 \\
C++/512z(512-bit) & double &
124.6 = 4.0 + 120.6 &
6.58E2 (x6.5) &
6.79E2 (x6.7) &
8.65E2 \\
\hline
C++/none(scalar) & float &
936.1 = 4.3 + 931.8 &
8.75E1 (x0.9) &
8.79E1 (x0.9) &
1.02E2 \\
C++/sse4(128-bit) & float&
228.9 = 3.9 + 225.0 &
3.58E2 (x3.6) &
3.64E2 (x3.6) &
4.30E2 \\
C++/avx2(256-bit) & float&
114.1 = 3.8 + 110.4 &
7.18E2 (x7.2) &
7.43E2 (x7.4) &
9.06E2 \\
C++/512y(256-bit) & float&
104.5 = 3.8 + 100.7 &
7.84E2 (x7.9) &
8.14E2 (x8.1) &
1.00E3 \\
C++/512z(512-bit) & float&
\hphantom{0}61.8 = 3.8 + \hphantom{0}58.0 &
1.33E3 (x13.3) &
1.41E3 (x14.1) &
1.77E3 \\
\hline
C++/none(scalar) & mixed &
986.0 = 4.3 + 981.6 &
8.31E1 (x0.8) &
8.35E1 (x0.8) &
9.98E1 \\
C++/sse4(128-bit) & mixed &
500.4 = 3.9 + 496.5 &
1.64E2 (x1.6) &
1.65E2 (x1.6) &
2.00E2 \\
C++/avx2(256-bit) & mixed &
220.5 = 3.8 + 216.7 &
3.72E2 (x3.7) &
3.78E2 (x3.8) &
4.55E2 \\
C++/512y(256-bit) & mixed &
195.6 = 3.7 + 191.8 &
4.19E2 (x4.2) &
4.27E2 (x4.3) &
5.21E2 \\
C++/512z(512-bit) & mixed &
118.5 = 3.8 + 114.7 &
6.92E2 (x6.9) &
7.15E2 (x7.2) &
8.97E2 \\
\hline
\end{tabular}
}\end{center}
\vspace*{-5mm}
\caption{
Processing times and throughputs 
for 
81952 \ggttggg\ weighted events.
One core
of Juwels Cluster login node jwlogin07
(Intel Gold 6148 CPUs), 
gcc11.2 builds.}
\label{tab-tputcppnewg}
\vspace*{#3}
\end{table}
}

\section{Introduction}

Computing architectures designed
for data parallelism,
such as CPUs with 
vector registers
and GPUs, are now ubiquitous 
in the computing resources 
used for the data processing of
High Energy Physics (HEP) experiments,
such as the
High Performance Computing (HPC)
centers available to the 
Large Hadron Collider (LHC) experiments
and the sites of the 
Worldwide LHC Computing Grid (WLCG).
The full compute power of
GPUs and vector CPUs, however,
is often underexploited in HEP processing,
partly
because the 
software is old
and was designed before these architectures
became mainstream, but also because
many HEP workflows 
involve a lot of stochastic branching and are therefore
intrinsically difficult to port to data parallel
paradigms,
one notable example being
detector simulation.
Monte Carlo (MC) matrix element generators,
conversely, are 
an ideal fit
to exploit these architectures.
This is because the calculation
of scattering amplitudes and matrix elements (MEs),
which is the computational bottleneck 
of these programs
for complex physics processes,
involves the repeated execution of the same functions
on different data items
(the various 
``events'' 
randomly generated by MC sampling),
and it is possible to achieve a
perfect lockstep processing
in its data parallel execution.

Our work on the reengineering of the 
Madgraph5\_aMC@NLO (MG5aMC)
event generator~\cite{bib:mg5amc} 
follows precisely 
this approach.
As described in our previous proceedings
of the vCHEP2021~\cite{bib:vchep2021}
and ICHEP2022~\cite{bib:ichep2022}
conferences, our new implementation
of the ME calculation 
in CUDA 
and vectorized C++ 
achieves lockstep processing
with 100\% branch efficiency 
on NVidia GPUs
and the maximum theoretically possible 
SIMD speedups (x8 and x16 in double and single
floating point precision for AVX512/zmm) 
on vector CPUs.
In this paper, 
we mainly document the results presented at the
ACAT2022 conference (October 2022),
where we had reported the process achieved
in the few months since ICHEP2022 (July 2022).
This includes in particular
some performance tests 
of the vectorized C++ implementation
using all cores of a CPU 
rather than a single CPU core,
some performance improvements
for the serial component
of the overall workflow, 
the implementation of a new
``mixed'' precision 
mode 
where both single and double
floating point precision 
are used for different parts 
of the ME calculation,
and the full integration
into the existing MadEvent framework
of the ME calculation
implemented using SYCL.
We also briefly mention 
a few new results
achieved since ACAT2022
at the time of 
writing (March 2023),
some of
which 
will be described 
in more detail 
in upcoming talks~\cite{bib:chep2023a,bib:chep2023b}.

As discussed 
in Ref.~\cite{bib:vchep2021},
our port of MG5aMC ME calculations
to CUDA and NVidia GPUs, which is based
on Feynman diagrams and helicity amplitudes,
represents a restart 
from scratch of previous 
efforts~\cite{bib:hagiwara1,bib:hagiwara2}
in this direction in 2009,
which unfortunately
were never integrated into a production
quality framework usable by HEP experiments.
Other approaches for porting
matrix element event generators to GPUs
have also been suggested. 
The development of 
MadFlow~\cite{bib:madflow1,bib:madflow2}
is another project
based on MG5aMC helicity amplitudes,
which however
is independent
from our work and differs from it because
it uses Python and the TensorFlow framework.
Furthermore,
a GPU port of ME calculations 
based on Berends-Giele recursion relations,
which was initially prototyped~\cite{bib:giele}
in 2010 as a way to achieve a better 
scalability with the number of external 
particles in the scattering process
than one based on Feynman diagrams
(polynomial rather than factorial),
has recently been implemented
into the new PEPPER simulation 
framework~\cite{bib:blockgen,bib:pepper}.

\section{Speeding up the serial component of
the MadEvent framework}

As we previously described 
in our ICHEP2022 proceedings~\cite{bib:ichep2022},
our strategy for delivering
to the LHC experiments
a software application that they can run
to generate samples of events,~with 
well-known user interfaces
and identical physics output
but at a fraction of 
current computational costs,
is based on injecting 
one of our new data-parallel
implementations~(in CUDA/C++ or SYCL)
of the ME calculation
into the existing MadEvent framework, 
replacing only the previous 
scalar 
Fortran implementation 
of the same ME calculation.
The ``outer shell'' of the 
MadEvent framework,
which is also implemented in Fortran,
takes care of all tasks other
than the ME calculation,
which we will collectively refer to
as the ``non-ME serial component'' of MadEvent:
this includes, amongst other things,
the generation of pseudo-random numbers,
their mapping to particle momenta
using a well defined sampling strategy
(based on the MadEvent single-diagram
enhancement multichannel 
algorithm~\cite{bib:madevent}),
the merging of multi-jet final states
(for instance using the 
so-called 
``MLM'' scheme~\cite{bib:mlm,bib:alwall08}),
the execution of the hit-or-miss
unweighting algorithm,
the calculation of cross sections
and the I/O intensive writing 
of LHE event data files.
While all these tasks
only account for a few percent
of the overall wall-clock time
when the Fortran serial MEs are used,
the situation changes dramatically
when the 
much
faster
(one to three orders 
of magnitude)
CUDA/C++ or SYCL
data-parallel MEs 
based on CPU vectorization or GPUs
are used,
as the MadEvent non-ME serial component
quickly becomes the bottleneck.

\tabtputcudanew{t}{-10mm}{-3mm}

In the results 
presented at ICHEP2022 
for the $\ggttgg$ process,
for instance, we had
reported that 
generating
90k weighted events
took 58.3 seconds overall 
(5.2s in the MadEvent non-ME serial component and
53.1s in the ME calculation,
see Table~2 in Ref.~\cite{bib:ichep2022})
using Fortran MEs,
but only 6.1 seconds overall
(5.7s non-ME and 0.36s MEs)
using double-precision CUDA MEs.
In other words,
the factor $\sim$200 speedup
in the ME calculation 
only led to
an overall speedup
by a factor $\sim$10:
this is the limit
predicted by Amdahl's law~\cite{bib:amdahl}
since the serial non-ME component
was originally 5.2s/58.3s,
i.e.~approximately 10\%
of the overall processing time.
Our new ACAT2022 
results
for the same process
are given in Table~\ref{tab-tputcudanew}:
the generation workflow
in the madevent executable
now takes 55.4 seconds overall 
(2.4s non-ME and 53.0s MEs)
using Fortran MEs,
but only 2.9 seconds overall
(2.6s non-ME and 0.35s MEs)
using double-precision CUDA MEs,
i.e.~a factor two faster
than in the ICHEP2022 results.
The difference between the two sets
of results is only in the 
MadEvent non-ME serial component,
which is now 
a factor two faster,
while the speed of the CUDA ME calculation
is essentially unchanged.
The overall speedup from Fortran to CUDA 
is now $\sim$20, as predicted
by Amdahl's law
since the serial component
was originally 2.4s/55.4s,
i.e.~approximately 5\%
of the overall processing time.

To explain this speed-up,
we recall~\cite{bib:ichep2022} that
the original 
MadEvent framework, which
was looping through individual events
and executing the full processing chain
(random sampling of momenta,
computing MEs, unweighting,
multi-jet merging etc.)
one event at a time,
had to be modified
to allow the data-parallel
calculation of MEs
on a large batch 
of events
at the same time:
this naturally led to the introduction
of large Fortran arrays 
to keep all relevant properties
of all the events 
in that batch.
In this particular case,
the speedup of the serial non-ME component
from 5.2s to 2.4s was obtained
by rationalizing the handling
of MLM multi-jet merging,
and in particular by moving
most of its processing
before the ME calculation, 
which made it possible to 
completely get rid of some
very large Fortran arrays
that had been introduced 
in the initial transformation
of MadEvent
from a single-event 
to a multi-event processing framework.

Speeding up 
the MadEvent serial non-ME component
is especially important when
offloading the ME calculation to a GPU,
but it 
remains relevant
when MEs are computed on vector CPUs.
For instance,
our new results for 
generating
80k $\ggttgg$ events
on an Intel Gold 6148 CPU,
which are given in Table~\ref{tab-tputcppnew},
show that the overall workflow
now takes 6.1 seconds
(1.8s non-ME and 4.3s MEs)
using our "512z" vectorization level
(AVX512 with zmm registers~\cite{bib:ichep2022}),
while at ICHEP2022
we had reported that the same
workflow on the same machine
took 
7.1 seconds
(2.5s non-ME and 4.5s MEs,
see Table~1 in Ref.~\cite{bib:ichep2022}).
Again, the difference between the two sets
of results mainly comes from
the MadEvent serial non-ME component,
but 
the effect 
of Amdahl's law
is less pronounced 
for C++
than for CUDA,
as the ME calculation is still
the bottleneck.

\tabtputcppnew{t}{-13mm}{-4mm}

\newcommand{\figbmkcuda}[3]{
\begin{figure}[#1]
\vspace*{#2}
\begin{center}
\includegraphics[width=0.70\textwidth,clip]{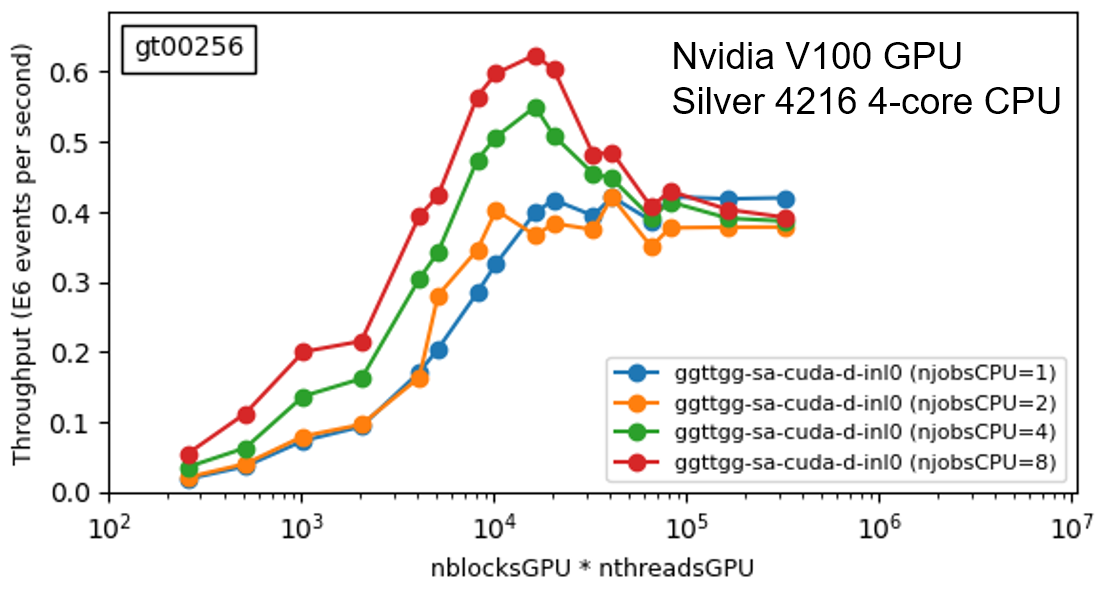}
\end{center}
\vspace*{-6mm}
\caption{Total combined throughput
for the $\ggttgg$ process
using 1, 2, 4 or 8 copies
of our standalone application
(see Ref.~\cite{bib:ichep2022}),
as a function of the CUDA grid size
(number of blocks per grid
times number of threads per 
block --- "gt00256" indicates that the latter is fixed to 256).
}
\label{fig:bmkcuda}
\vspace*{#3}
\end{figure}
}

\figbmkcuda{b}{-3mm}{-9mm}

\newcommand{\figbmkcpp}[3]{
\begin{figure}[#1]
\vspace*{#2}
\begin{center}
\includegraphics[width=0.88\textwidth,clip]{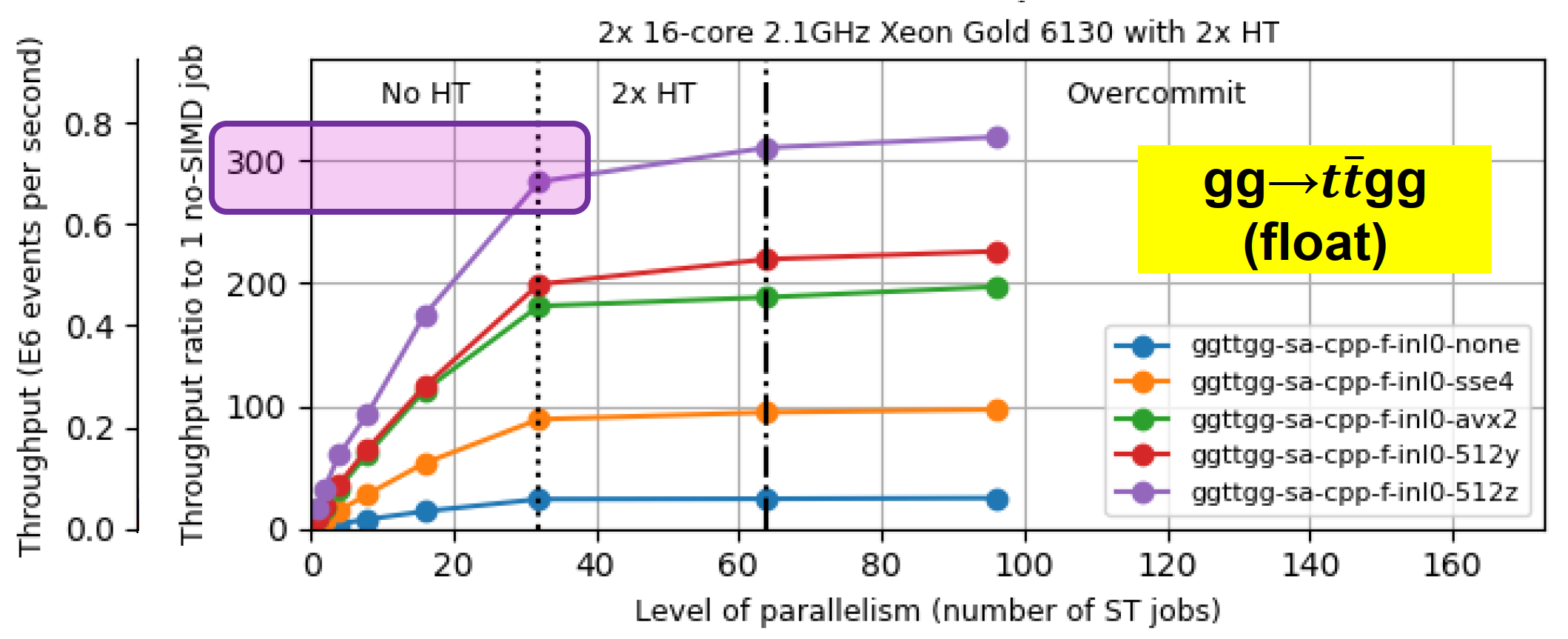}
\end{center}
\vspace*{-6mm}
\caption{Total combined throughput
for the $\ggttgg$ process
as a function of~the~number~of 
copies of our 
single-threaded
standalone application,
in our 
five C++ 
vectorization scenarios.
The y-axis~represents
the ratio of the achieved throughput
to a reference 
with no vectorization 
and a single CPU process.
For reference, the range 
of values of the absolute 
throughputs is also shown.
The x-axis represents 
the number of simultaneous
processes
launched on the test node,
which includes 
32 physical CPU cores 
with Hyper-Threading enabled
(64 logical 
cores in total).
}
\label{fig:bmkcpp}
\vspace*{#3}
\end{figure}
}

While this speed-up
in MadEvent
is already an important achievement, 
we think that this is just the first step
and that there is still much potential
for further performance
improvements.
Further rationalizations of the 
use of large Fortran arrays
may still be possible.
In addition, we are investigating ways
to speed up 
the MadEvent serial non-ME component
by parallelizing it at least in part.
One idea, for instance, 
is to offload to the GPU
(or vectorize on the CPU)
some parts of the computation,
such as the mapping from random 
numbers to momenta 
in the sampling algorithm,
or the unweighting process.
Another possible approach, which
represents
a truly
heterogeneous processing scenario,
would consist in running several
copies of the madevent application
in parallel 
on different CPU threads,
while sharing the GPU amongst 
them
for the ME calculation.
In addition to speeding up
the MadEvent 
non-ME component by 
parallelizing it
amongst different CPU cores,
another 
advantage of this approach
is that
it could 
allow a decrease 
in the RAM footprint 
of each madevent process on the CPU
(which is problematic 
as discussed in Ref.~\cite{bib:ichep2022}),
as it should be possible to
achieve the same 
overall occupancy of the GPU
while decreasing
the number of events
computed in parallel 
by a single madevent process, 
i.e.~its CUDA grid size.
The results of a 
preliminary test 
relevant to this approach
are displayed in Fig.~\ref{fig:bmkcuda},
which shows the variation 
of the combined ME throughput 
achievable from a single NVidia V100 GPU
when this is shared by up to 8
processes running in parallel
on different CPU threads.
The notable effect that we were
hoping to see, 
and which is indeed achieved,
is that the throughput curve
moves to the left as the number
of CPU processes increases, while still
reaching the same combined
throughput plateau at the end:
this means that the maximum
GPU throughput may be reached
by running many CPU applications
with smaller CUDA grid sizes,
rather than a single application
with a very large grid size.
Another
positive result,
which however
we were not anticipating and
will deserve more in-depth analysis,
is the fact that the maximum
combined GPU throughput
actually increases by almost 50\%
when launching kernels from
different CPU threads.
It should be stressed that this
plot, which was obtained using the
infrastructure developed for the
HEP-SCORE benchmarking project~\cite{bib:bmk},
refers to the ``standalone'' 
application~\cite{bib:ichep2022}
where the ME calculation is not yet
integrated in the full MadEvent workflow:
in the future, we plan to repeat
similar studies using the full MadEvent 
workflows, which would represent
a more realistic test of a 
production-like heterogeneous scenario.

\figbmkcpp{t}{-15mm}{-3mm}

\section{Further performance tests and improvements in the ME calculation}

In parallel to our efforts to
understand and 
speed up
the MadEvent serial 
non-ME component,
we have also continued to pursue
further improvements and analyses
of the ME calculations.

To start with, based on the
same benchmarking infrastructure
that we used to produce 
Fig.~\ref{fig:bmkcuda}
for the CUDA back-end,
we analysed the performance
of our vectorized C++ back-end
when several CPU cores are used.
This differs from the results 
that we presented in our previous
papers as well as in 
Table~\ref{tab-tputcppnew} above,
which all refer to a single CPU core.
The results of this test are given 
in Fig.~\ref{fig:bmkcpp}.
One effect that is immediately
visible is that the AVX512/zmm 
throughputs (purple line)
continue to be significantly 
faster than the AVX512/ymm (red)
and AVX2 (green) throughputs
even when many cores are used,
but not by a factor two.
This may be due to a clock slowdown,
but we have not verified it.
With respect to non-vectorized
throughput on a single core,
the overall speedup of AVX512/zmm
with 32 processes
(the number of physical cores
in these two Intel Gold 6130 CPUs)
using single floating point precision
is approximately 300,
compared to a theoretical
maximum of 512 (32 times 16),
which seems quite satisfactory.

\tabtputcudanewg{b}{-3mm}{-9mm}

\tabtputcppnewg{t}{-15mm}{-3mm}

Another progress
in the CUDA/C++ back-end
has been the addition 
of a ``mixed''~floating precision mode,
where Feynman diagrams are
computed in double precision,
while the ``color algebra''
part of the ME calculation
is done
in single precision.
The rationale for this~approach
is that 
floats provide
approximately a factor two
speedup over
doubles
both in vectorized C++
(because twice as many floats as doubles
fit into the same vector register)
and in CUDA (because typical
NVidia data center cards
have twice 
as many FLOPs
for FP32 as 
for FP64),
but single precision does not
provide enough numerical precision
for the Feynman diagram 
part of the ME calculation.
The idea was to test whether single
precision could at least be used
for the ``color algebra'':
our tests confirmed 
that the same
cross sections could be obtained
within~$\sim$10$^{-5}$
in this case,
which seems enough.
Our throughput results 
for the $\ggttggg$ process
are shown in Table~\ref{tab-tputcudanewg} 
for CUDA and 
Table~\ref{tab-tputcppnewg} 
for vectorised C++.
While encouraging, these results
are still preliminary and
we plan to pursue further tests
of this approach.

\section{SYCL-based developments and C++ compiler studies}

\newcommand{\figsycl}[3]{
\begin{figure}[#1]
\vspace*{#2}
\begin{center}
\includegraphics[width=0.72\textwidth,clip]{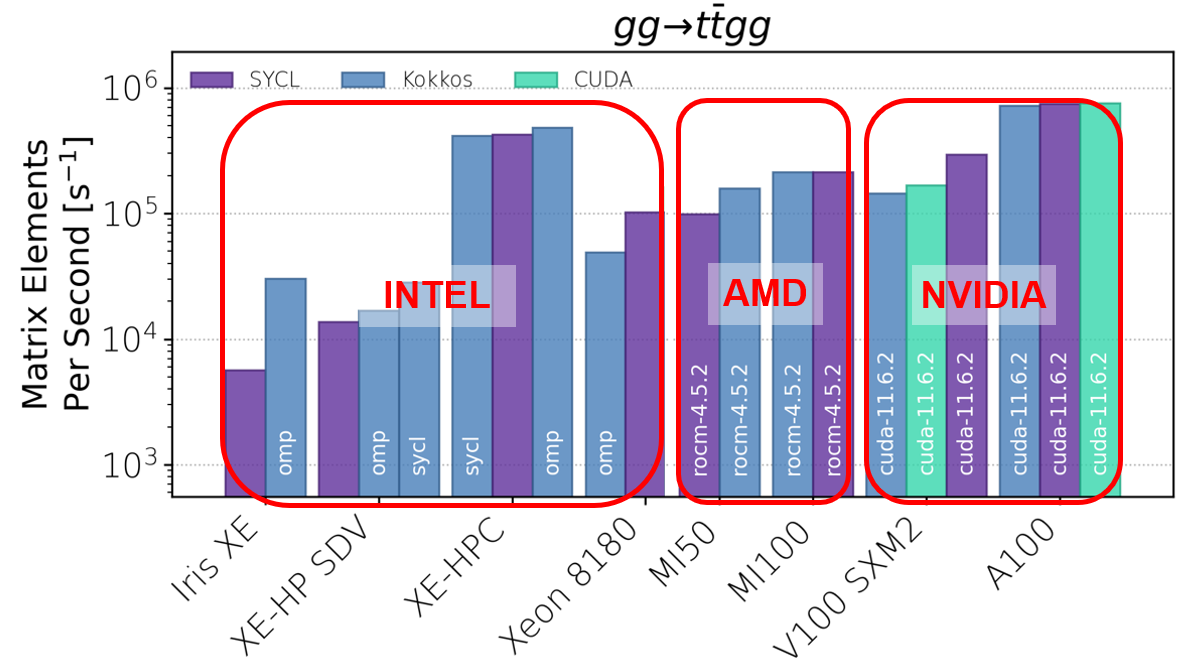}
\end{center}
\vspace*{-6mm}
\caption{
Comparison
of the CUDA, Kokkos and SYCL
ME engines
for $\ggttgg$ 
on many GPUs,
using the standalone application
(with optimal GPU grid sizes 
at the throughput plateau).
``Xe-HP SDV'' is a Software Development Vehicle
for functional testing only,
currently used at Argonne and at other customer sites 
to prepare their code for future Intel data centre GPUs.
``XE-HPC'' is an early implementation of the Aurora GPU.
The throughput achieved
on a full Xeon 8180 CPU 
using SYCL and Kokkos 
multi-threading
is also shown for reference.
}
\label{fig:sycl}
\vspace*{#3}
\end{figure}
}
\figsycl{b}{-3mm}{-9mm}

\newcommand{\figol}[3]{
\begin{figure}[#1]
\vspace*{#2}
\begin{center}
\includegraphics[width=0.86\textwidth,clip]{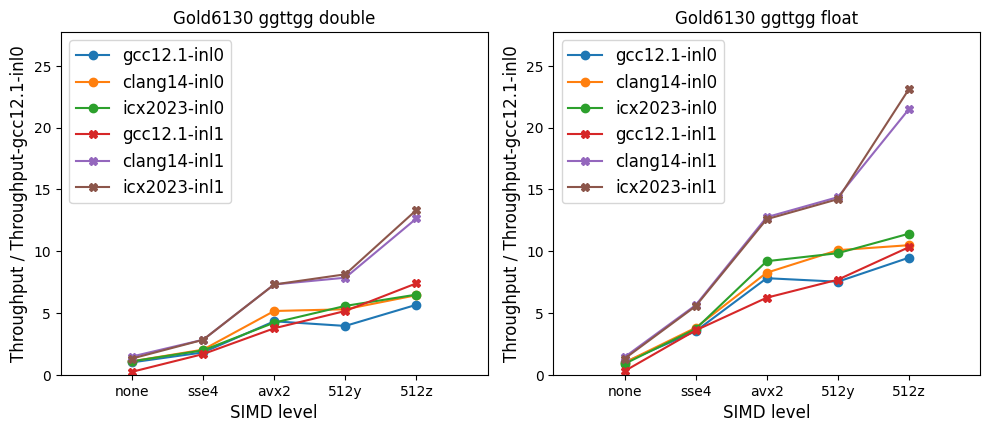}
\end{center}
\vspace*{-6mm}
\caption{
Comparison~\cite{bib:openlab}
of $\ggttgg$ throughputs
for different C++ builds,
using~five vectorization levels,
three compilers
(gcc12.1; clang14.0;
icx2023, based on clang16.0)
and two inlining approaches.
The inl0 default build uses 
no inlining,
the inl1 experimental build 
(which mimics
some features of link time optimization)
forces the inlining
of helicity amplitude functions.
Rather than absolute throughputs,
all data points in 
each plot
(left for double,
right for single precision)
represent the ratios 
to the throughput
gcc12.1 inl0 build
with no SIMD.
}
\label{fig:ol}
\vspace*{#3}
\end{figure}
}

While all tables and plots presented 
so far in this paper refer
to our original CUDA/C++ implementation,
significant progress has also been 
achieved on various fronts in our
parallel implementations using
performance
portability frameworks.
Most recently, this work has 
focused on the SYCL implementation,
while the developments using Kokkos have slowed down 
and those based
on Alpaka have stopped.
As noted in Ref.~\cite{bib:ichep2022},
the main interest
of these APIs
is that a single code base,
with a few back-end-specific
customizations,
may be executed 
on many architectures,
including GPUs from different 
vendors such as NVidia,
AMD and Intel.
This is shown 
in Fig.~\ref{fig:sycl},
which compares the performances
of our CUDA, SYCL and Kokkos
implementations on
different systems;
compared to
previous 
results~\cite{bib:ichep2022},~this~ACAT2022~plot 
is~interesting
because it 
also includes
results on Intel XE-HPC,
which is an early implementation
of the Aurora GPU.
A notable
achievement 
reported
at ACAT2022
is that 
the SYCL implementation 
of the ME calculation is now also
fully integrated into MadEvent,
which means for instance that we 
are able to produce cross-sections
and LHE event data files
by offloading the ME calculation
to AMD or Intel GPUs,
rather than using 
the Fortran CPU implementation.

A more recent 
development,
which started well 
after ACAT2022, 
is that a vectorized SYCL implementation for CPU
has also been prototyped.
Preliminary tests indicate
that this achieves a promising
performance, with throughputs
which sometimes exceed those 
of the gcc builds 
of the CUDA/C++ implementation:
while this is not yet understood and will require further studies,
it is likely that this may be due at least in part to the fact that the SYCL implementation is built
using the clang-based 
icx Intel compiler.
As shown in Fig.~\ref{fig:ol},
in fact,
which presents 
a recent~\cite{bib:openlab}
performance comparison
between many builds
of the CUDA/C++ implementation
using different
C++ compilers,
we have observed that 
the performance of icx builds 
is almost the same as that 
of clang builds,
which can be significantly
better than that of gcc
builds in some cases
(more than a factor 2 faster
with AVX512/zmm vectorization
and agressive inlining);
these results are 
however
preliminary and will need
more in-depth analysis.
It is also 
interesting to note that,
while our CUDA/C++ implementation
of vectorization is based on gcc and clang
compiler vector extensions,
our SYCL version
uses the sycl::vec type, 
which is 
itself implemented as a wrapper
over clang vector extensions:
in other words, compiler vector
extensions are ultimately used
for CPU vectorization
in both of our
CUDA/C++ and SYCL
implementations.

\figol{t}{-15mm}{-3mm}

\section{Outlook: towards a first alpha release}

Finally,
the most important progress
we 
achieved since ACAT2022
is that we 
completed
the implementation of
the event-by-event 
random choice of
leading colors~and~helicities~in~LHE~files.
This was the last missing piece
before we could provide 
in the CUDA/C++ MadEvent framework
the full set of features needed
by the LHC experiments
for unweighted event generation.
This functionality is now
essentially complete,
but we are still performing 
some final tests,
also to understand its
impact on performance;
in particular, 
this feature
introduces a minor level
of stochastic branching
in the ME workflow,
degrading lockstep
processing both on GPUs
and on vector CPUs
(it is possible that 
this effect is already
visible in Fig.~\ref{fig:ol},
which was prepared using 
this more recent code base).
We are now working towards
repackaging our work to provide
a first alpha release of our work 
for the experiments, 
which we plan 
to achieve
during Q2 2023.

\section*{Acknowledgements}

This research used resources 
of the Argonne Leadership Computing Facility, 
which is a DOE Office of Science User Facility 
under contract DE-AC02-06CH11357, 
and of
the Joint Laboratory for System Evaluation (JLSE)
at Argonne National Laboratory.
We also gratefully acknowledge the use 
of computing resources
at CINECA under ISCRA-C project MG5A100
and at the J\"ulich Supercomputing Centre 
at Forschungszentrum J\"ulich
under PRACE-DEV-2022D01-022.


\section*{References}
\begin{thebibliography}{9}

\bibitem{bib:ichep2022}
A. Valassi et al.,
PoS(ICHEP2022)212 (2022). 
\doi{10.22323/1.414.0212}

\bibitem{bib:mg5amc} 
J. Alwall et al.,
JHEP07(2014)079.
\doi{10.1007/JHEP07(2014)079}

\bibitem{bib:vchep2021}
A. Valassi et al., 
EPJ Web of Conferences 251, 03045 (2021). 
\doi{10.1051/epjconf/202125103045}

\bibitem{bib:chep2023a}
S. Hageb\"ock, 
CHEP2023 presentation,
\url{https://indico.jlab.org/event/459/contributions/11829/}

\bibitem{bib:chep2023b}
Z. Wettersten, 
CHEP2023 presentation,
\url{https://indico.jlab.org/event/459/contributions/11850/}

\bibitem{bib:hagiwara1}
K. Hagiwara et al., 
Eur. Phys. J. C \textbf{66} (2010) 477.
\doi{10.1140/epjc/s10052-010-1276-8}

\bibitem{bib:hagiwara2}
K. Hagiwara et al., 
Eur. Phys. J. C \textbf{70} (2010) 513.
\doi{10.1140/epjc/s10052-010-1465-5}

\bibitem{bib:madflow1}
S. Carrazza et al.,
EPJ Web of Conferences 251, 03022 (2021).
\doi{10.1051/epjconf/202125103022}

\bibitem{bib:madflow2}
S. Carrazza et al.,
Eur. Phys. J. C \textbf{81} (2021) 656.
\doi{10.1140/epjc/s10052-021-09443-8}

\bibitem{bib:giele}
W. Giele et al.,
Eur. Phys. J. C \textbf{71} (2011) 1703.
\doi{10.1140/epjc/s10052-011-1703-5}

\bibitem{bib:blockgen}
E. Bothmann et al.,
SciPost Phys. Codebases 3 (2022).
\doi{10.21468/SciPostPhysCodeb.3}
    
\bibitem{bib:pepper}
E. Bothmann et al.,
\arxiv{2311.06198}.
 
\bibitem{bib:madevent}
F. Maltoni, T. Stelzer, 
JHEP02(2003)027.
\doi{10.1088/1126-6708/2003/02/027}

\bibitem{bib:mlm}
M. L. Mangano et al., 
Nucl. Phys. B \textbf{632} (2002) 343.
\doi{10.1016/S0550-3213(02)00249-3}

\bibitem{bib:alwall08}
J. Alwall et al., 
Eur. Phys. J. C \textbf{53} (2008) 473. 
\doi{10.1140/epjc/s10052-007-0490-5}

\bibitem{bib:amdahl}
G. M. Amdahl, 
Computer \textbf{46} (2013) 38.
\doi{10.1109/MC.2013.418}

\bibitem{bib:bmk}
D. Giordano et al., 
Comput Softw Big Sci 5, 28 (2021). \doi{10.1007/s41781-021-00074-y}

\bibitem{bib:openlab}
A. Valassi,
CERN Openlab workshop (March 2023),
\url{https://indico.cern.ch/event/1225408/contributions/5243830/}



 
\end{thebibliography}
\end{document}